\documentclass[12pt]{iopart}

\usepackage[T1]{fontenc}
\usepackage{graphicx}
\usepackage{iopams}
\usepackage{color}
\usepackage{hyperref}
\graphicspath{{./figures/}{./}}

\newcommand{\Dp}{\ensuremath{D_\mathrm{p}}}
\newcommand{\Dobs}{\ensuremath{D_\mathrm{obs}}}
\newcommand{\Deff}{\ensuremath{D_\mathrm{eff}}}
\newcommand{\ee}{\mathrm{e}}

\newcommand{\changes}[1]{{#1}}
\newcommand{\changesf}[1]{{#1}}

\begin{document}

\title{Scaling study of diffusion in dynamic crowded spaces}

\author{Harry~Bendekgey$^{1,2}$, Greg~Huber$^1$, 
David~Yllanes$^{1,3,4,5}$}

\address{$^1$ Chan Zuckerberg Biohub --- SF, 499 Illinois Street, San Francisco, California 94158, USA}
\address{$^2$ Department of Computer Science, University of California, Irvine, California 92697, USA}
\address{$^3$ Fundación ARAID, Diputación General de Aragón, 50018 Zaragoza, Spain}
\address{$^4$ Instituto de Biocomputaci\'{o}n y F\'{i}sica de Sistemas Complejos (BIFI), 50018 Zaragoza, Spain}
\address{$^5$ Zaragoza Scientific Center for Advanced Modeling, 50018 Zaragoza, Spain}
\eads{\mailto{david.yllanes@bifi.es}}

% Add these email addresses: 
%   gerghuber@gmail.com and hbendekg@uci.edu
% Orchid ID for Huber: 
%  0000-0001-8565-3067

\date{\today}

\begin{abstract}
\changesf{We formulate a scaling theory for the long-time diffusive motion
in a space occluded by a high density of moving obstacles in
dimensions 1, 2 and 3}.  Our tracers diffuse anomalously over many decades in time, before reaching a diffusive steady state with an effective diffusion
constant $D_\mathrm{eff}$, which depends on the obstacle diffusivity and density.
The scaling of $D_\mathrm{eff}$, above and below a critical regime,
is characterized by two independent critical
parameters: the conductivity exponent $\mu$, also found in models with frozen
obstacles, and an exponent $\psi$, which quantifies the effect of obstacle
diffusivity.
\end{abstract}
\submitto{\jpa}

%\maketitle 

%%%%%%%%%%%%%%%%%%%%%%%%%%%%%%%%%%%%%%%%%%%%%%%%%%%%%%%%%%%%%%%%%%%%%%%%%%%

\newcommand{\sigobs}{\sigma_{\mathrm{obs}}}
\newcommand{\sigp}{\sigma_{\mathrm{p}}}
\newcommand{\nhat}{\ensuremath{\hat{n}}}
\newcommand{\nhatc}{\ensuremath{\hat{n}_\mathrm{c}}}

\section{Introduction} 
Brownian motion in disordered media has long been the focus of much theoretical
and experimental work~\cite{Havlin1987, Bouchaud1990,Avraham2000,Benichou2014}.  A particularly important
application has more recently emerged, due to the progress in imaging and
microscopy techniques, namely, transport inside the cell (see,
e.g.,~\cite{Schwille1999,Smith1999,Dayel1999,Seisenberger2001} for some
pioneering studies or~\cite{Novak2009,Saxton2012,Hofling2013,Mogre2020} for
more recent overviews).  Indeed, it is now possible to track the movement of
single molecules or other small particles inside living cells, with sizes
ranging from small proteins to viruses, RNA molecules or ribosomes.  One
generally considers the mean square displacement $\langle \Delta r^2(t)\rangle$
(MSD, the average of the squared distance traveled by the particle) and tries
to fit it to a law of the form $\langle \Delta r^2(t)\rangle \propto t^\alpha$.
Some experiments in prokaryotic cytoplasms~\cite{Bakshi2011,Coquel2013} have
found evidence for this behavior with $\alpha=1$, characteristic of the
classical Brownian (or diffusive) motion~\cite{Berg1983}. Many other works,
however, have found signals of subdiffusive (i.e., $\alpha<1$)
transport~\cite{Golding2006,Weber2010}.  In eukaryotic cells, where the
intracellular environment is considerably more complex~\cite{Brown2020}, heterogeneous~\cite{Ghosh2016} and
crowded, the range of observed behavior is even
wider~\cite{Bronstein2009,Jeon2011,Tabei2013}.

On the theoretical front, in addition to the general problem of (sub)diffusion, attention 
has been paid to issues such as first-passage times~\cite{Condamin2007,Benichou2014,Guerin2016},
driven~\cite{Illien2013,Benichou2014b,Illien2018} and active~\cite{Mogre2018,Rizkallah2022}
tracers, different particle shapes and boundary conditions~\cite{Benichou2018,Klett2021,Alexandre2022}.
A celebrated framework to generate anomalous diffusion is the continuum-percolation or ``Swiss-cheese''
model~\cite{Halperin1985}. In it, a large number of interpenetrating
obstacles, usually disks or spheres, are randomly distributed throughout the
system (see Figure~\ref{toy}).  The behavior of tracer particles is controlled
by the obstacle density. For low concentration, the long-time behavior is
diffusive.  When the percolation threshold~\cite{Stauffer1994,Bollobas2006}  is
reached, however, the system undergoes a localization
transition~\cite{Saxton1994,Hofling2006,Kammerer2008,Bauer2010,Spanner2011}.  The
Swiss-cheese model is successful in generating (transient) anomalous diffusion
in a crowded environment. From the point of view of cellular transport,
however, it falls short in one crucial respect: 
cellular interiors are constantly rearranging 
themselves. This
shortcoming can be addressed by considering dynamical obstacles~\cite{Tremmel2003,Schmit2009, Dorsaz2010, Berry2014,Polanowski2016, Nandigrami2017}.
In this case, the localization transition disappears and
normal diffusion is reached at \emph{any} obstacle density for long times~\cite{Saxton1987}.
This regime is usually described with the aid of a theory presented by Nakazato and Kitahara~\cite{Nakazato1980},
%This regime is usually described with the aid of a theory presented by Nakatazo and Kitahara~\cite{Nakazato1980},
which is exact in the low- and high-density limits. For intermediate densities, however,
the theory of \cite{Nakazato1980} offers good quantitative agreement only if 
the obstacles move relatively fast~\cite{Tahir-Kheli1983, Beijeren1985}.

Here we take an alternative approach to the problem by considering
a wide range of timescales, from very slow obstacle diffusion
to the faster regimes studied by previous authors\changes{, with the goal
of finding universal critical scaling behavior (see, e.g.,~\cite{cardy:96,zinn-justin:05,amit:05,pelissetto:02}).} Using
extensive numerical simulations, we compute
an effective diffusion constant, \Deff, for a wide range of 
obstacle concentrations and diffusivities in one, two and 
three spatial dimensions. Even though no 
localization transition is observed, we find that the 
critical percolation point still holds a special importance.
In particular, the value
of \Deff\ is controlled by two critical exponents:
the conductivity $\mu$~\cite{Adler1985}, which quantifies
the approach to the critical density, and an exponent
$\psi$, giving the scaling with the obstacle diffusivity.

\section{Model and parameters}
\begin{figure}[tb]
\centering
\includegraphics[width=0.7\columnwidth]{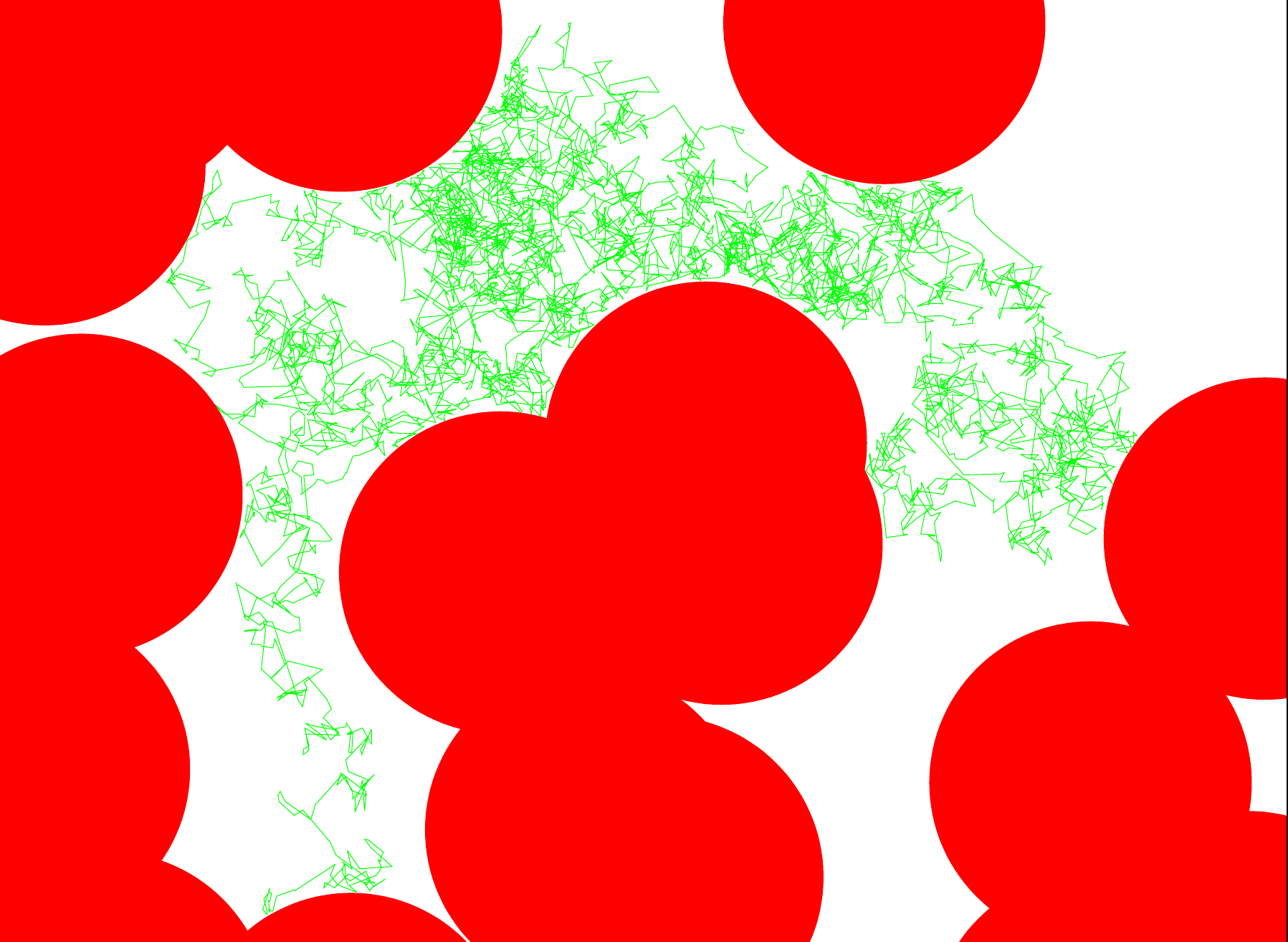}
\caption{A particle traces a discrete-time random walk in two dimensions  in
green, blocked by frozen obstacles. At each time interval it draws a proposal
for its new location from a centered Gaussian distribution. If offered a new
position inside a red obstacle, the particle stays put for that timestep and
attempts a new move at the following timestep. This arrangement of overlapping
circular obstacles is commonly referred to as the Swiss-cheese model~\cite{Halperin1985}
in the case of immobile obstacles. Here, we 
consider the obstacles to be also Brownian particles, diffusing
independently from one another.}
\label{toy}
\end{figure}

In our model, which we implement in
continuous spaces in spatial dimension $d=1,2$ and $3$, a volumeless particle
undergoes a discrete-time random walk and encounters obstacles that block its
path. The obstacles are uniformly-placed possibly-overlapping $d$-balls, an
arrangement that has been called the Swiss-cheese model \cite{Halperin1985},
see Fig.~\ref{toy}. At each timestep, our tracer particle takes a step in a random direction (chosen isotropically) of length extracted from a Gaussian distribution
with standard deviation $\sigp$. We set $\sigp=1$ to define the unit length
in the system. 

 \changes{The precise form of the interaction between walkers and obstacles seems, at first glance, to be a delicate issue. On second glance, however, the expectation is that it will affect the short-time dynamics rather than the scaling properties of the asymptotic behavior of the tracers. This assertion can be rigorously motivated by the theory and experimental results on critical phenomena and the renormalization group~\cite{cardy:96,amit:05,zinn-justin:05}. Briefly stated, around
a critical point, one can identify a small number of scaling variables whose evolution can encode the behavior of a complex system with many degrees of freedom. One 
then finds that microscopic details cease to matter and, instead, systems can be classified into a small number of universality classes, characterized only by
symmetry considerations and other global features, such as the number of spatial 
dimensions. Hence, all the systems in the same universality class can be understood through the study of their simplest representative. The most celebrated example of this universality is the Ising model, naively a toy representation of magnetic interactions, but which quantitatively explains experimental results not only in ferromagnetic systems, but also for the liquid-gas transition, and for binary fluids~\cite{pelissetto:02}.

We are motivated to employ this approach here by the known results for
diffusion in a static disordered medium. In this case, as we shall explain
below, the system experiences a phase transition and, hence, universal scaling
behavior as it approaches a percolation limit. This scaling is characterized
by a critical \emph{conductivity} exponent $\mu$, so called because it can
account for the behavior of conductivity of conductor/superconductor mixtures
as well as for that of random walks in a static, crowded environment~\cite{Grassberger1999}.

With these considerations in mind, we shall define simple interaction rules.
If we then find power-law dependences for the diffusion constant, the 
resulting critical exponents ---or, in other words, our results for the long-time behavior  of the diffusivity--- will be applicable to systems with more realistic
interactions. Therefore, we} simply have the particle stop if its proposed new location would
be inside one of the obstacles.  To avoid situations where the particle can
jump over obstacles, we fix the obstacle radius as $R = 10 \gg \sigp$. 
The obstacles themselves also move, ignoring interactions and always drawing
their discrete steps from a Gaussian distribution with standard deviation
$\sigobs \leq \sigp$. To deal with situations where the void pocket inhabited
by a particle becomes entirely squeezed, obstacles can ``step on'' particles,
in which case the particle stops all motion until the obstacle moves off of it.

We are interested in how well the particle can explore the empty space, as
measured by the mean square displacement after $t$ timesteps,
\begin{equation}\label{eq:D}
\langle \Delta r^2(t)\rangle = \langle [\boldsymbol r(t+t_0) - \boldsymbol
r(t_0)]^2 \rangle. 
\end{equation}
As discussed in the introduction, when the MSD 
grows linearly with time, we say the motion is diffusive 
and characterize it by a diffusion constant $D$, defined
through the equation $\langle \Delta
r^2(t)\rangle = 2dDt$, where $d$ is the spatial dimension.

The first control parameter in this model is the dimensionless obstacle
density, $\hat{n} = nR^d$, where $n$ is the number density of obstacle centers.
In previous works, in which the obstacles are frozen, at low
obstacle densities the MSD is governed by three different behaviors at varying
time scales \cite{Hofling2006,Bauer2010}. These time
scales have also been observed empirically in biological systems
\cite{Saxton2007}.

In these cases, at microscopic time scales, particles do not encounter
obstacles and diffuse freely. This gives us another
characteristic parameter: the short-term diffusivity of the tracers,
$\Dp = {\sigp^2}/{2d}$. If there were no obstacles at all,
we would recover $\langle \Delta r^2(t)\rangle = 2d\Dp t$. 

At intermediate time scales, the system experiences subdiffusive
motion, meaning $\langle \Delta r^2(t)\rangle \sim t^\alpha, \alpha < 1$.
Finally, for long times, the central limit theorem causes the MSD to revert to
linear growth, $\langle \Delta r^2(t)\rangle \sim 2d\Deff t$ with $\Deff \ll \Dp$
\cite{Hofling2013}.

In frozen models, this third window of diffusive motion recedes as $\hat{n}$
approaches a critical value, $\hat{n}_\mathrm{c}$, at which point it
disappears entirely.  This localization transition takes place at the percolation point
$\hat n_\mathrm{c}$ of the obstacles, when an infinite cluster
partitions space into finite, disconnected
pockets~\cite{Stauffer1994,Bollobas2006}.  The critical
scaling of \Deff\ is controlled by a conductivity exponent
$\mu$~\cite{Adler1985,Grassberger1999,Hofling2006}
\begin{equation}\label{eq:mu}
\Deff \propto (\nhatc - \nhat)^{\mu},
\end{equation}
with $D_{\mathrm{eff}} = 0$ in the supercritical regime $\hat{n} \geq
\nhatc$.  For $d=2$ and $3$, the critical density has been calculated to be
$\nhatc = 0.359$~\cite{Quintanilla2007,Mertens2012} and $0.839$~\cite{Yi2012}, respectively.
We also see that $\nhatc = 0$ trivially in $d=1$, because localization occurs once a
single frozen obstacle appears. 

In our model with moving obstacles, as has been
observed~\cite{Tremmel2003,Berry2014}, particles never become localized because
the obstacles themselves will eventually move out of the way. We expect, thus,
to find these three time windows at all obstacle densities, matching the
biological observations.  As we shall demonstrate, however, \nhatc\ still
carries a special significance, so we continue to refer to $\nhat <\nhatc$
($\nhat >\nhatc$) as the subcritical (supercritical) regime.

The second control parameter is the diffusivity of the obstacles, $\Dobs$.
This will generally be much lower than \Dp\  (otherwise, with our rules,
the tracers' movement would become enslaved to the obstacles).
To optimize our simulations, we only move the obstacles every
10 time steps, giving us $D_{\mathrm{obs}} =
{\sigma_{\mathrm{obs}}^2}/{20d}$. As $\Dobs \to 0$ we recover the
frozen models~\footnote{We have checked this explicitly by
computing fits to~(\ref{eq:mu}) and confirming
that our $\mu$ is compatible with the values in the literature.}.

\begin{figure}[tb]
\centering
\includegraphics[width=0.7\columnwidth]{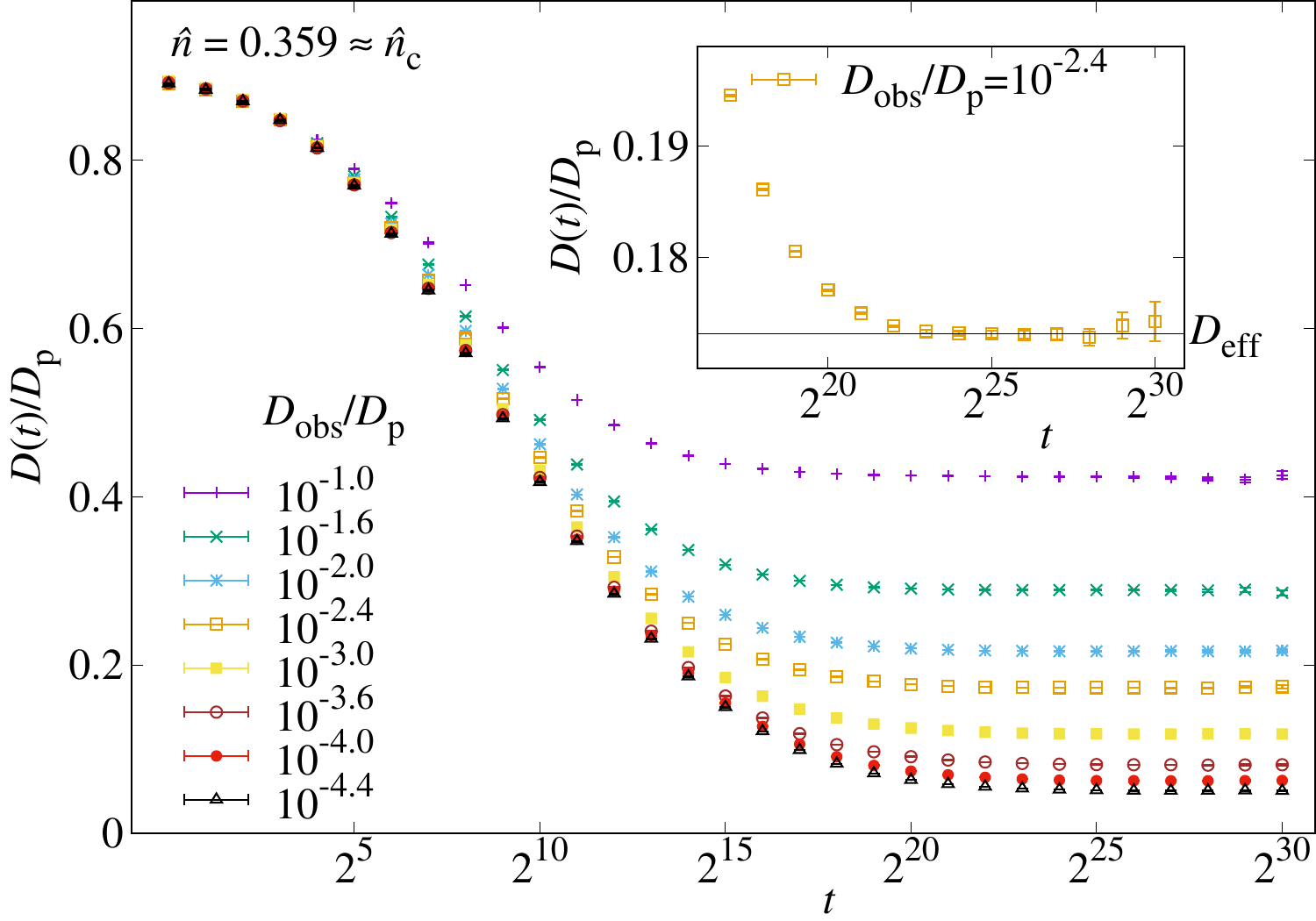}
\caption{Time-dependent diffusion coefficient $D(t)$, eq.~(\ref{eq:Dt}), in $d=2$ at
approximately critical obstacle density for various obstacle diffusivities.
Diffusivities are shown in units of the characteristic particle diffusivity $D_\mathrm{p}$.
In this representation, diffusive motion is characterized by a constant $D(t)$.
Such a regime is always reached for long times in our model, no matter the obstacle density,
defining an effective diffusion constant \Deff. The inset shows a closeup 
of the approach to this steady-state  behavior for  $\Dobs/\Dp = 10^{-2.4}$.
All curves in this plot are averaged over $24$ independent simulations, each
with $360$ tracer particles.
}
\label{dt}
\end{figure}

\section{Results in \boldmath $d=2$} We use 
a simulation box of linear size $L=1000$ with periodic boundary conditions.
Given the annealed nature of our disorder, this is more than enough to avoid
finite-size effects (see~\ref{finite}).  Since we are interested in
the asymptotic behavior, we need to follow our tracers for very long  times in
order to obtain good statistics; we use $1.2\times 10^9$ time steps. The  MSD
for each tracer is calculated according to eq.~(\ref{eq:D}), for values of $t$
chosen as powers of 2 and with $t_0$ ranging over multiples of $10^6$.  Thus
we averaged the distance traveled by a single particle across a shifting time
window, with each $10^6$ steps as a different starting time. We gave the system
a $10^7$ timestep ``burn-in'' period, in which data was not collected to allow
the system to converge to its stationary distribution. We further improved our
estimate of $\langle\Delta r^2(t)\rangle$ by launching 360 independent tracers
in a single simulation, and ran 24 separate simulations for each pair of
$\sigobs,\nhat$ values, so that the configuration and motion of obstacles can
vary from system to system.

We describe the motion of the particle over time with a time-dependent
diffusion coefficient,
\begin{equation}\label{eq:Dt}
D(t) = \frac{\langle \Delta r^2(t)\rangle}{2dt\phi}\ .
\end{equation}
This definition is similar to that of \cite{Bauer2010}, with one key
difference: because some particles are
trapped under obstacles at any given moment and therefore not moving, time must
be rescaled by the proportion of particles that are free, or
$\phi=\ee^{-\pi\nhat}$ in $d=2$
\footnote{Equivalently, $\phi = e^{-2\nhat}$ in $d=1$ and $\phi =
e^{-\frac{4}{3}\pi\nhat}$ in $d=3$.}. A similar argument
was made in~\cite{Novak2011,Novak2011b}. This allows us to recover
$D(t)\approx \Dp$ for very small $t$, and also allows our results to 
approach the proper values in the limit of $\Dobs\to0$.
Figure~\ref{dt} displays the evolution of $D(t)$ over time  
for simulations at $\nhat\approx\nhatc$ (similar plots
in the sub- and supercritical regime are included in~\ref{subsup}).
We are interested in the asymptotic value $\Deff = \lim_{t\to\infty}D(t)$, which,
as discussed above, can be reached for any \nhat.

\begin{figure}[tb]
\centering
\begin{minipage}[t]{.49\linewidth}
\includegraphics[width=\columnwidth]{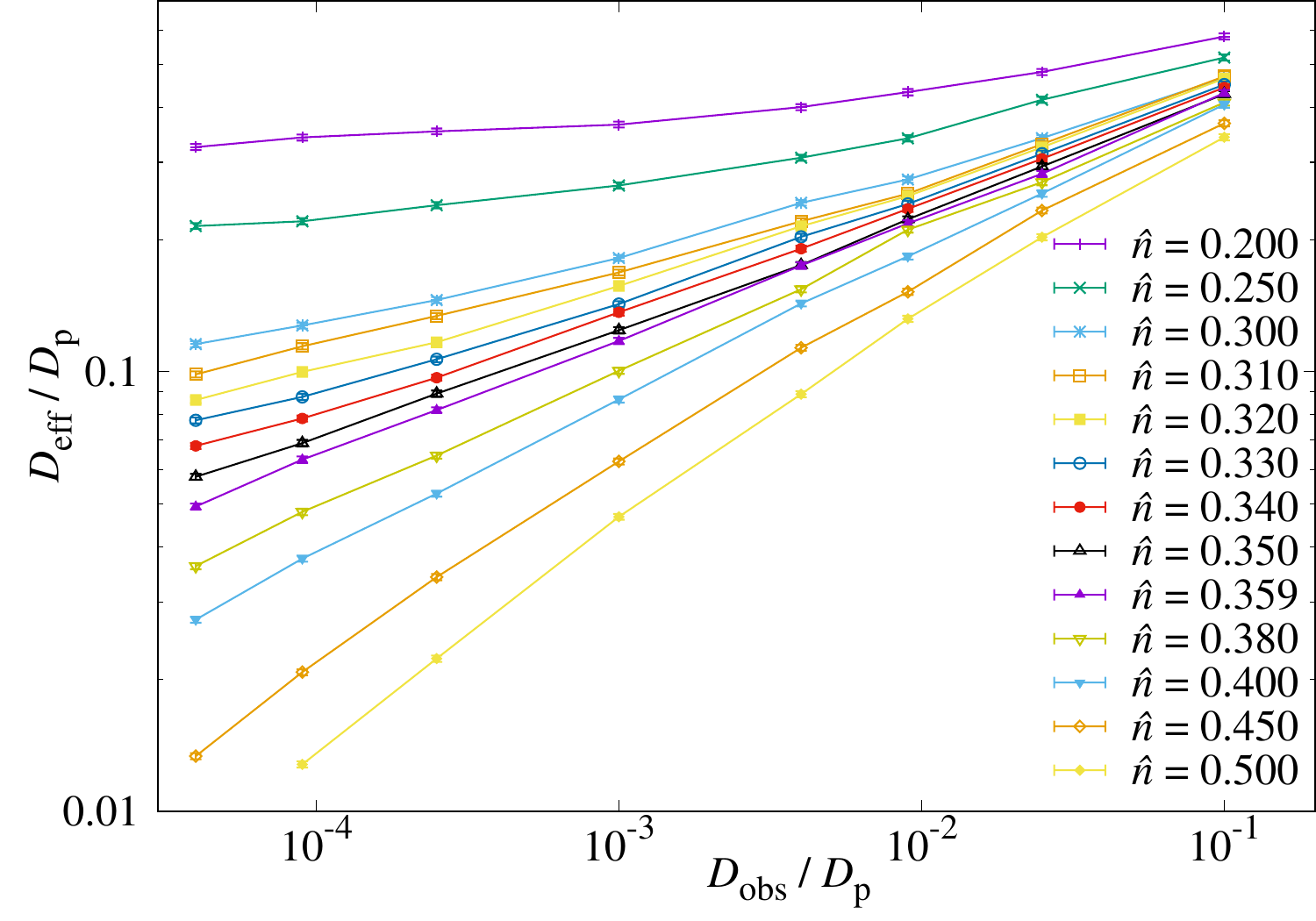}
\end{minipage}
\begin{minipage}[t]{.49\linewidth}
\includegraphics[width=\columnwidth]{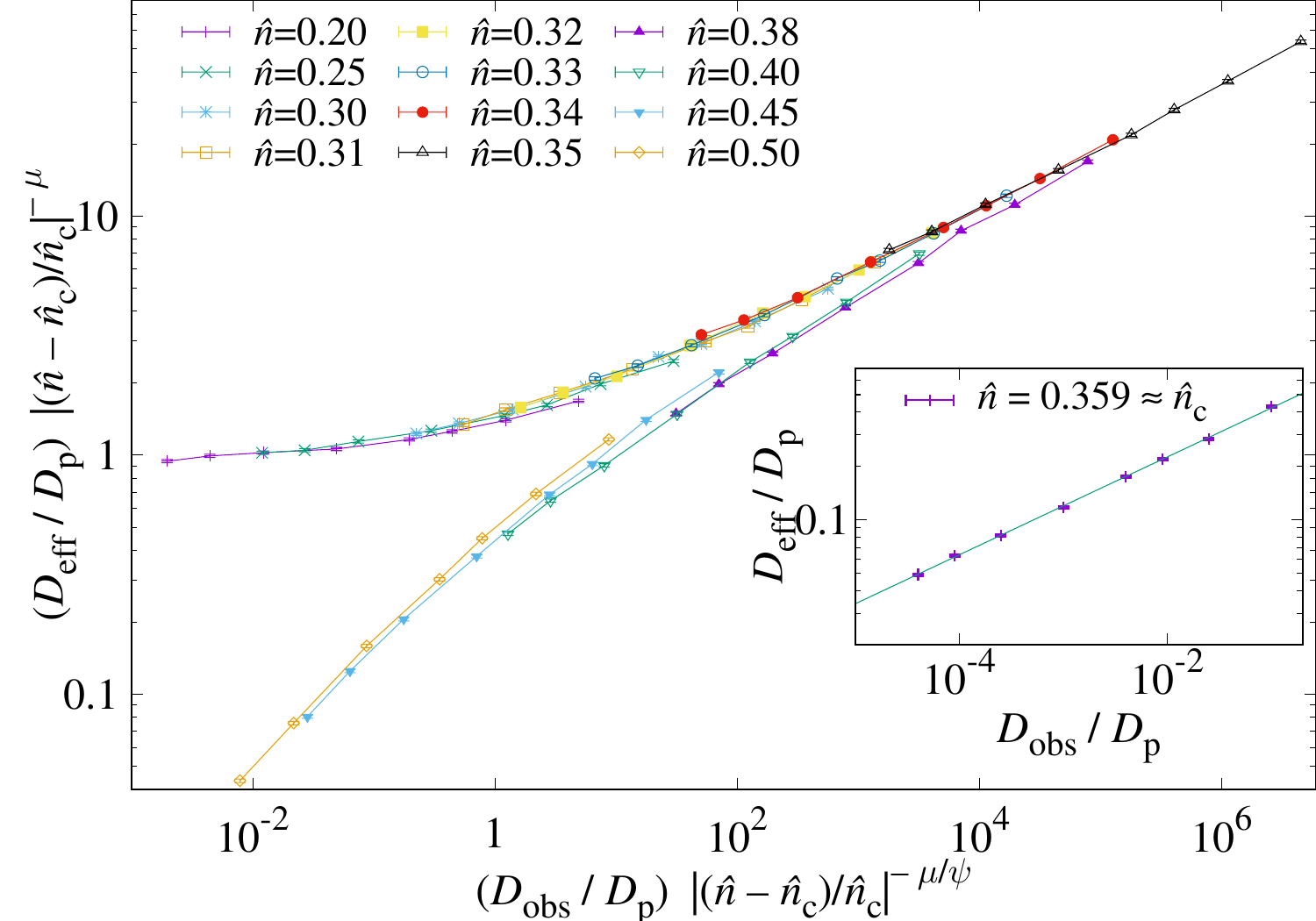}
\end{minipage}
\caption{\emph{Left:} Effective diffusion coefficient against the diffusion
coefficient of the obstacles for many obstacle densities in $d=2$.
The percolation density is $\hat n_\mathrm{c} \approx 0.359$.
\emph{Right:} Scaling behavior in $d=2$  of \Deff\
around the critical point, $\nhatc \approx 0.359$, according
to eq.~(\ref{eq:scaling}). We 
take our value of the conductivity exponent, $\mu = 1.31$,
from~\cite{Grassberger1999}.  The dynamic exponent $\psi$
is computed with a critical fit to $\Deff \propto \Dobs^\psi$
at $\hat n = \hat n_\mathrm{c}$ (inset), yielding
$\psi= 0.274(2)$ with $\chi^2/\mathrm{d.o.f.} = 6.42/6$.
Using these exponents, the data collapse in the main panel, with
separate branches for the supercritical and subcritical regimes, is achieved
without adjustable parameters.
\label{fig:2d}}
\end{figure}

We estimate $\Deff$ by fitting $\langle\Delta r^2(t)\rangle$ to
a constant for $t\geq2^{29}$. Error bars are computed
with a bootstrap method~\cite{Young2012}. The  resulting
values are plotted in Figure~\ref{fig:2d}--left. For $\nhat < \nhatc = 0.359$, as $\Dobs \to 0$,
$\Deff$ must eventually plateau at the values found by the frozen-obstacle
models. On the other hand, for $\nhat \geq \nhatc$, $\lim_{\Dobs \to 0}\Deff =
0$.  As a consequence, in our log-log plots, curves for $\nhat<\nhatc$ are concave and 
those for $\nhat>\nhatc$ are convex. Remarkably, at precisely \nhatc,
\Deff\ follows a power law
\begin{equation}\label{eq:psi}
\Deff \propto  \Dobs^\psi,\qquad \nhat = \nhatc.
\end{equation}
A fit to eq.~(\ref{eq:psi}) (see inset to Figure~\ref{fig:2d})
yields $\psi = 0.274(2)$ with and excellent 
goodness-of-fit value of
$\chi^2/\mathrm{d.o.f.} = 6.42/6$
[d.o.f. = degrees of freedom].
\changes{Our numerical finding of a power-law behavior for \Deff\
suggests that our model could exhibit critical scaling even in the case
of moving obstacles, with the addition of a critical exponent 
$\psi$ that encodes the dependence on obstacle diffusivity just 
as $\mu$ encodes the effect of obstacle density.}

Combining eq.~(\ref{eq:psi}) with eq.~(\ref{eq:mu}), we 
can formulate the following ansatz for the critical scaling 
in our model:
\begin{eqnarray}\label{eq:scaling}
\Deff/\Dp \simeq |\epsilon|^\mu\ g_{\pm}\bigl[(\Dobs/\Dp)\ |\epsilon|^{-\mu/\psi}\bigr], 
\end{eqnarray}
where $\epsilon = (\nhat -\nhatc)/{\nhatc}$ and $g_{\pm}(x)$
are the scaling functions that describe the behavior in the sub- and
supercritical regimes. Asymptotically, we expect $g_\pm(x) \sim x^\psi$ as $x
\to \infty$, and $g_-(x) \sim \mathrm{const.}$ as $x \to 0$. Thus, in the
subcritical regime, we recover eq.~(\ref{eq:mu})
as in studies with frozen obstacles. Using 
$\mu=1.3100(11)$~\cite{Grassberger1999} and our previously
computed $\psi$, eq.~(\ref{eq:scaling})  collapses
all our data without any adjustable parameters, see Figure~\ref{fig:2d}--right.

\begin{figure}[t]
\begin{minipage}[t]{.49\linewidth}
\includegraphics[width=\columnwidth]{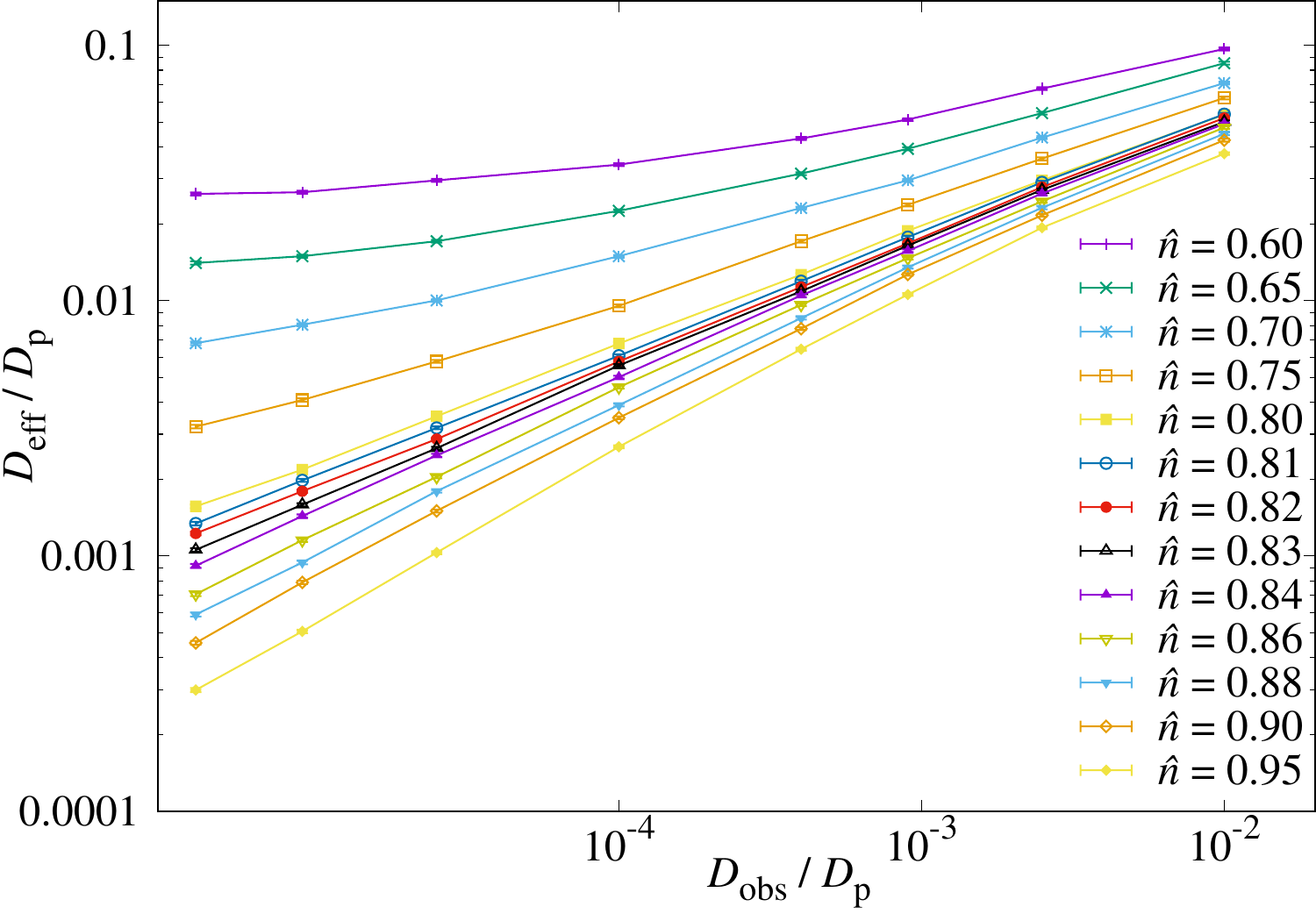}
\end{minipage}
\begin{minipage}[t]{.49\linewidth}
\includegraphics[width=\columnwidth]{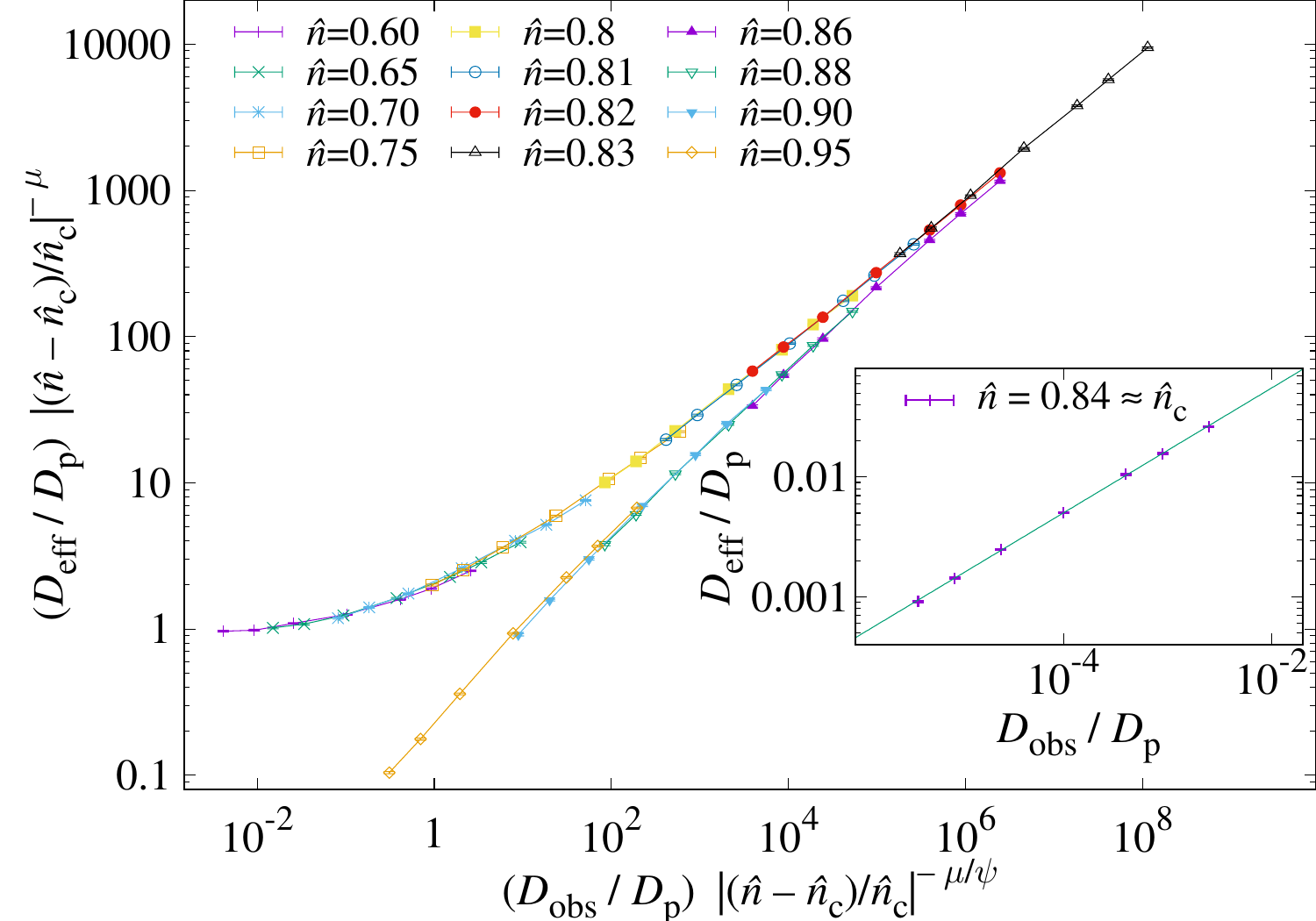}
\end{minipage}
\caption{As in Figure~\ref{fig:2d}, but in $d=3$. Now
$\hat n_\mathrm{c} \approx 0.84$ and $\mu\approx2.88$~\cite{Machta1985,Hofling2006}.
Our critical fit results in $\psi=0.520(3)$, 
with $\chi^2/\mathrm{d.o.f.} = 7.87/5$.
\label{fig:3d}}
\end{figure}

\section{Results in \boldmath $d=3$ } We ran analogous simulations in $d=3$.
To keep the wall-clock time of our runs under control, we reduced the
simulation box to $L=200$, and launched 320 tracers in each system. To avoid
finite-size effects, we further slowed the obstacles by only moving them once
every 100 steps.  We follow the same procedure as in $d=2$ to
compute $\psi=0.520(3)$, see Figure~\ref{fig:3d}

One difference between $d=2$ and $d=3$ is that in the latter case,
 universality
breaks down between lattice and continuum percolation models, so $\mu \neq
\mu_{\mathrm{lat}}$ \cite{Halperin1985}. There is, hence, some
disagreement on the value of $\mu$ in $d=3$. Following the 
analysis in~\cite{Machta1985,Hofling2006} we have used $\mu=2.88$,
which produces an excellent collapse of our data (see Figure~\ref{fig:3d}).

\section{Results in \boldmath $d=1$} The one-dimensional case has considerable
theoretical interest. Now tracers become localized at $\nhatc=0$, so there is
no subcritical regime.  For our non-frozen model, particles are not localized
because they can follow the obstacles as they move, or let an obstacle pass
over them.  We see, therefore, that $\psi=0$ trivially.

\begin{figure}[t]
\centering
\includegraphics[width=0.7\columnwidth]{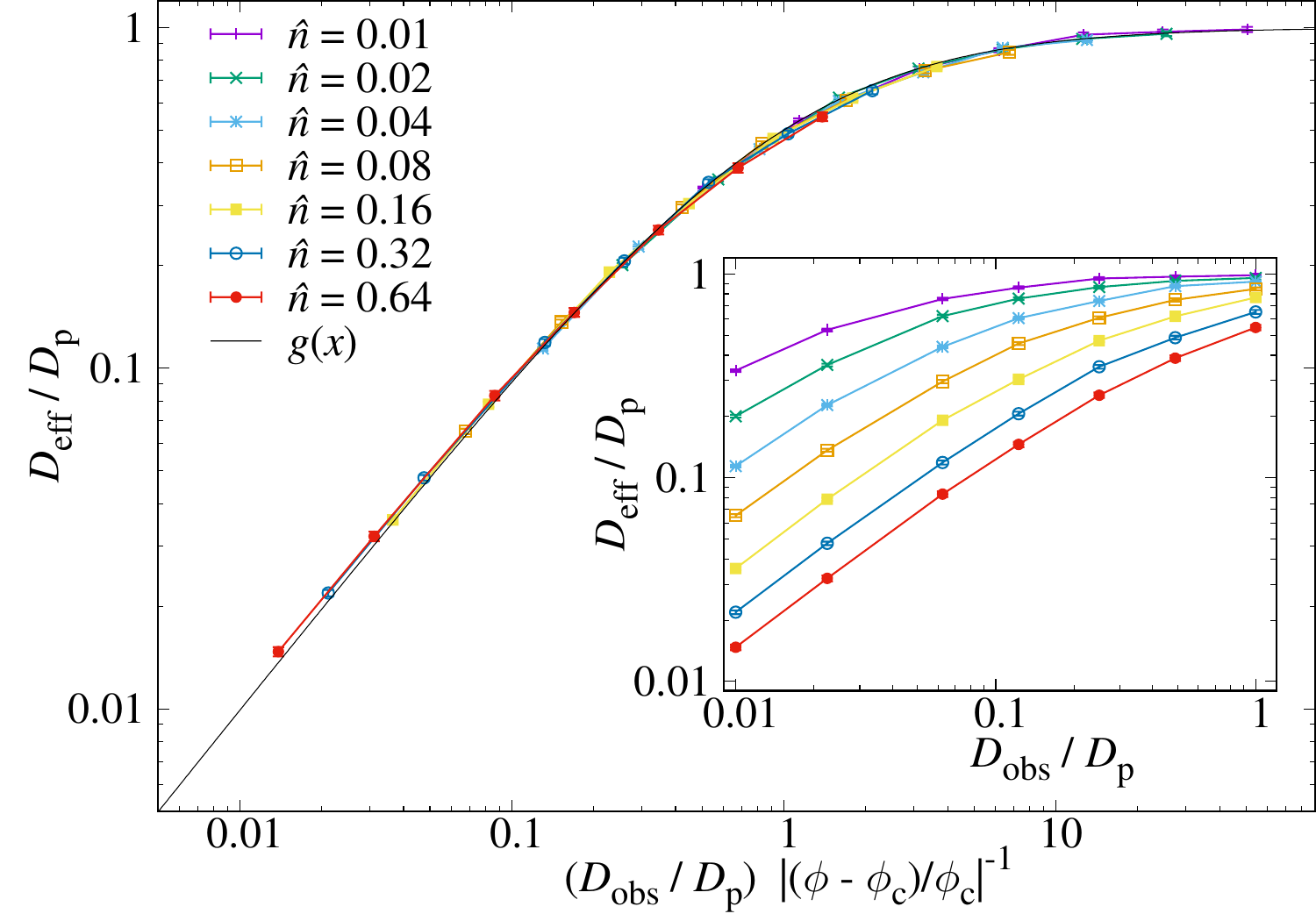}
\caption{Effective diffusion constant in $d=1$. 
We plot our \Deff\ rescaled according 
to eq.~(\ref{eq:ansatz}).
Notice that now all densities $\hat n>0$
are supercritical. The inset shows the unscaled
data.}
\label{1dscaled}
\end{figure}

In order to understand the scaling of \Deff\ in $d=1$, we begin by considering
several limits. First, when $\nhat\to0$, $\Deff\to\Dp$. Second, 
when $\Dobs\to0$ we approach the frozen limit with $\Deff=0$.
Finally, $\lim_{\nhat\to\infty} \Deff = \Dobs$. The latter is because, 
for dense enough obstacles, the tracer's motion is entirely
governed by how quickly the obstacles let it through.
With these in mind, and remembering that $\phi=\ee^{-2\nhat}$, we propose the following ansatz:
$\Deff = {\Dobs\Dp} / [{\Dp(1-\phi)+\Dobs}]$.
This is equivalent to using a different scaling variable
than was used in $d=2$ and $d=3$, namely $\epsilon_\phi = (\phi-\phi_\mathrm
{c})/\phi_\mathrm{c} = \phi - 1$, and writing 

\begin{equation}\label{eq:ansatz}
\Deff / \Dp = g\bigl[(\Dobs / \Dp)\ |\epsilon_\phi|^{-1}\bigr], \qquad g(x) = \frac{x}{1+x}.
\end{equation}
Figure~\ref{1dscaled} shows 
that eq.~(\ref{eq:ansatz}) is an excellent 
match to our data over a very wide range of $\phi$.

\section{Conclusions} We have performed a comprehensive 
investigation of diffusion in the void space of a dynamic
version of  the Swiss-cheese model in dimensions one through three.
We have characterized the critical point governing the 
anomalous diffusion using two exponents, one of which is the 
traditional conductivity exponent $\mu$, and the other of which 
we call $\psi$. The latter exponent describes scaling of 
the effective diffusivity at precisely the critical 
obstacle density. For dimensions one through three
we find $\psi=0$, $\psi=0.274(2)$ and $\psi=0.520(3)$,
respectively.

Together, $\mu$ and $\psi$ can be used
to encode the quantitative behavior of the system
for a wide range of obstacle densities and 
diffusivities. \changes{The physics of critical phenomena and the renormalization
group suggests that this scaling, which covers several orders of magnitude
in the mobility of the obstacles, will be universal, so our results}
should be directly relevant to the interpretation 
of cellular transport and other experiments
on crowded spaces.

Our scaling theory may have implications for a very different
kind of system: the low-temperature phase of spin ice~\cite{Harris1997, Castelnovo2012}. These compounds
produce emergent magnetic monopoles~\cite{Castelnovo2008} whose
dynamical behavior leads to an anomalous noise spectrum~\cite{Dusad2019,Samarakoon2022}, characterized by a scaling as $1/f^{1.5}$, different
from the $1/f^2$ of a paramagnetic system. For the case of 
Dy$_2$Ti$_2$O$_7$, this spectrum has been explained through
the presence of a subset of spins with a near-vanishing transverse 
field, whose flips are forbidden, which severely constrains
the movement of the monopoles~\cite{Hallen2022}. The resulting percolation picture maps onto the problem of diffusion of a particle in a space
of static obstacles. Other spin-ice compounds, such as those based on holmium, can generate very different transverse-field distributions.  
In these cases, the timescale of the slow spins can no longer be taken to infinity~\cite{Hallen2022,Tomasello2019} and has to be explicitly considered
in the scaling. The scaling theory presented here for a dynamic crowded space can provide such a quantitative description.

\section*{Acknowledgments}
We are grateful to Mark Veillette
and Le Yan for ideas and discussions at an early stage of the work, and to Roderich Moessner 
for bringing ferromagnetic spin ices to our attention.
H.B., G.H. and D.Y. were supported by the CZ Biohub -- SF.
G.H. received additional support from CZI Science.
This work has also been supported in part by Ministerio de Ciencia, Innovación y Universidades (Spain), Agencia Estatal de Investigación (AEI, Spain, 10.13039/501100011033), and European Regional Development Fund (ERDF, A way of making Europe) through Grant PID2022-136374NB-C21.
Our simulations were carried out at CZ Biohub -- SF and 
on the Cierzo supercomputer at BIFI-ZCAM (Universidad de Zaragoza).

\appendix
  
\section{Finite-Size Effects}

\label{finite}
\begin{figure}[b]\centering
\includegraphics[width=0.7\columnwidth]{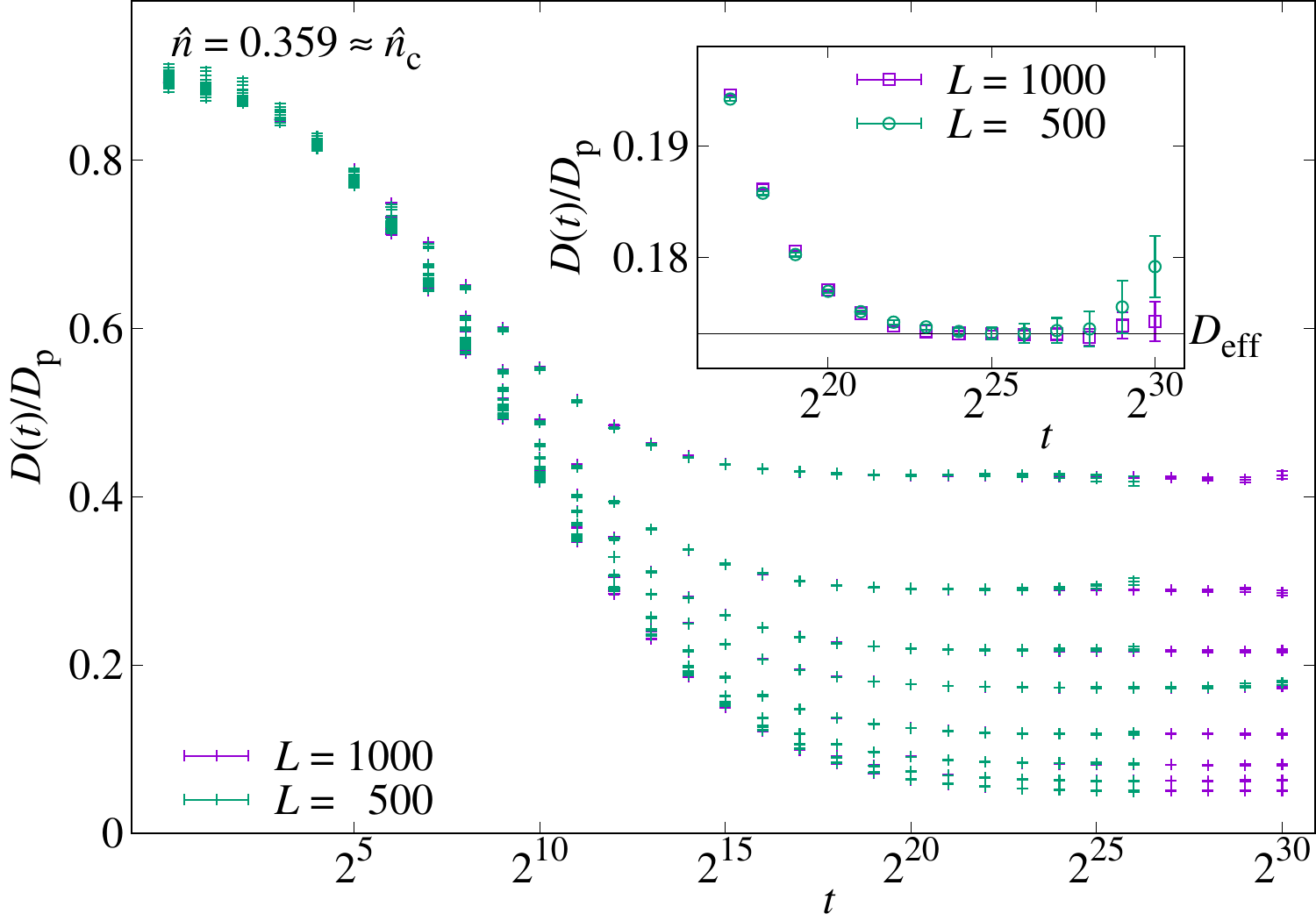}
\caption{Same as Figure~\ref{dt}, but adding 
data for a smaller system size ($L=500$). Since the curves for
$L=1000$ and $L=500$ are identical for the whole $t$ range, we have only
plotted most of the small simulations up to $t=2^{26}$ for visualization
purposes. The exception is $\Dobs/\Dp=10^{-2.4}$, enlarged in the inset.
The $L=500$ curve are noisier for long $t$
where there are few $t_0$ to compute the MSD.
\label{dtS}}
\end{figure}
Our model does not suffer from strong finite-size effects, probably thanks to the
annealed nature of the disorder. Therefore, our simulations for $L=1000$
should be representative of the large-$L$ limit.
We have, nevertheless, explicitly checked against finite-size effects by repeating
our $\nhat=\nhatc$ simulations for a smaller system size ($L=500$)
and verifying that we get the same $D(t)$ curve as in Figure~\ref{dt}.
This is shown in Figure~\ref{dtS}. For all $D_\mathrm{obs}$ the resulting
$\Deff$ is the same in the $L=500$ system, but the data is noisier.
Moreover, ultimately the strongest proof for the absence of 
finite-size effects is the fact that the power-law fits in Figures~\ref{fig:2d}
and~\ref{fig:3d} satisfy a $\chi^2$ test.

\section{Time evolution in the sub- and supercritical regimes}
\label{subsup}
\begin{figure}
\begin{minipage}[t]{.49\linewidth}
\includegraphics[width=1.0\columnwidth]{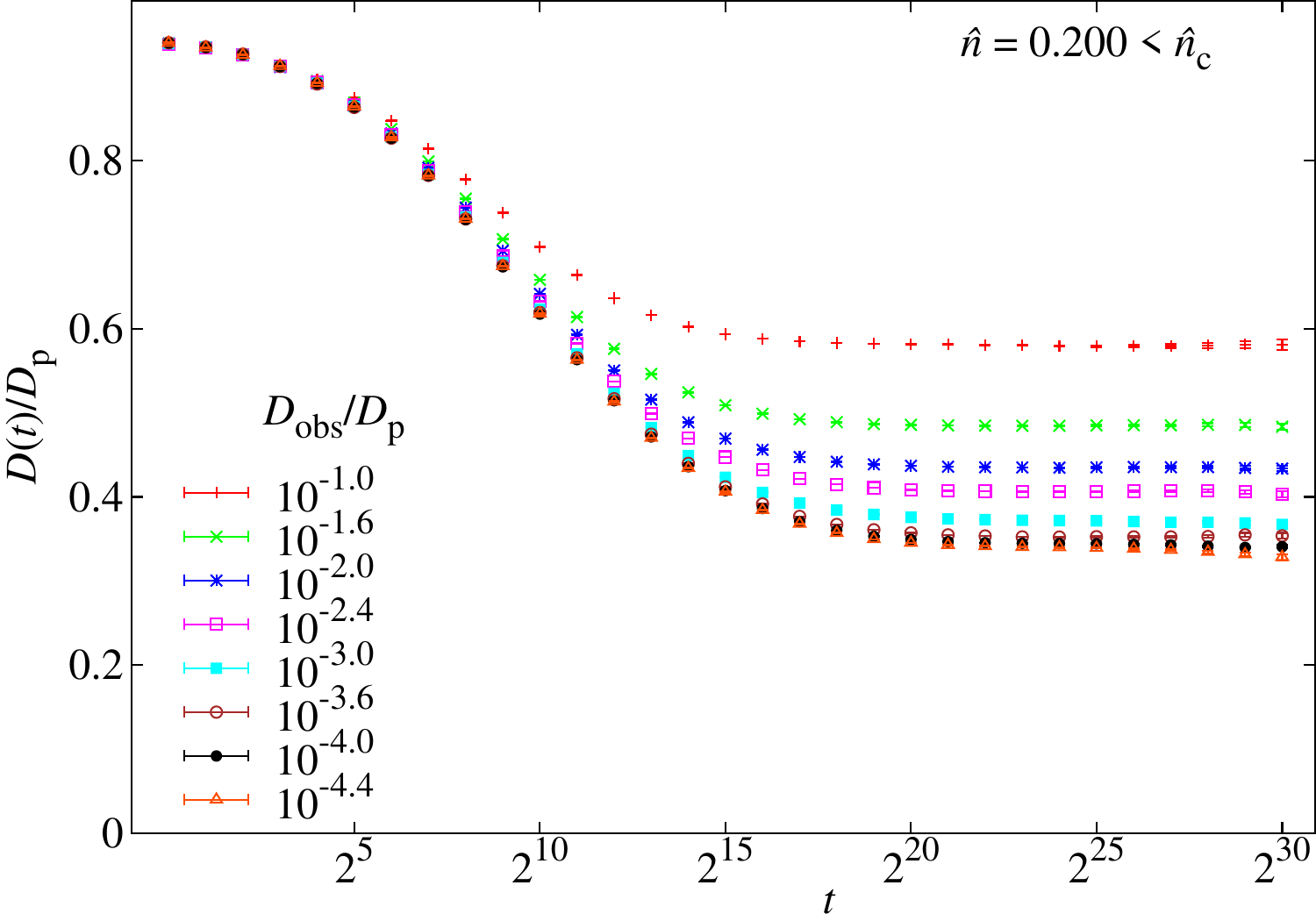}
\end{minipage}
\begin{minipage}[t]{.49\linewidth}
\includegraphics[width=1.0\columnwidth]{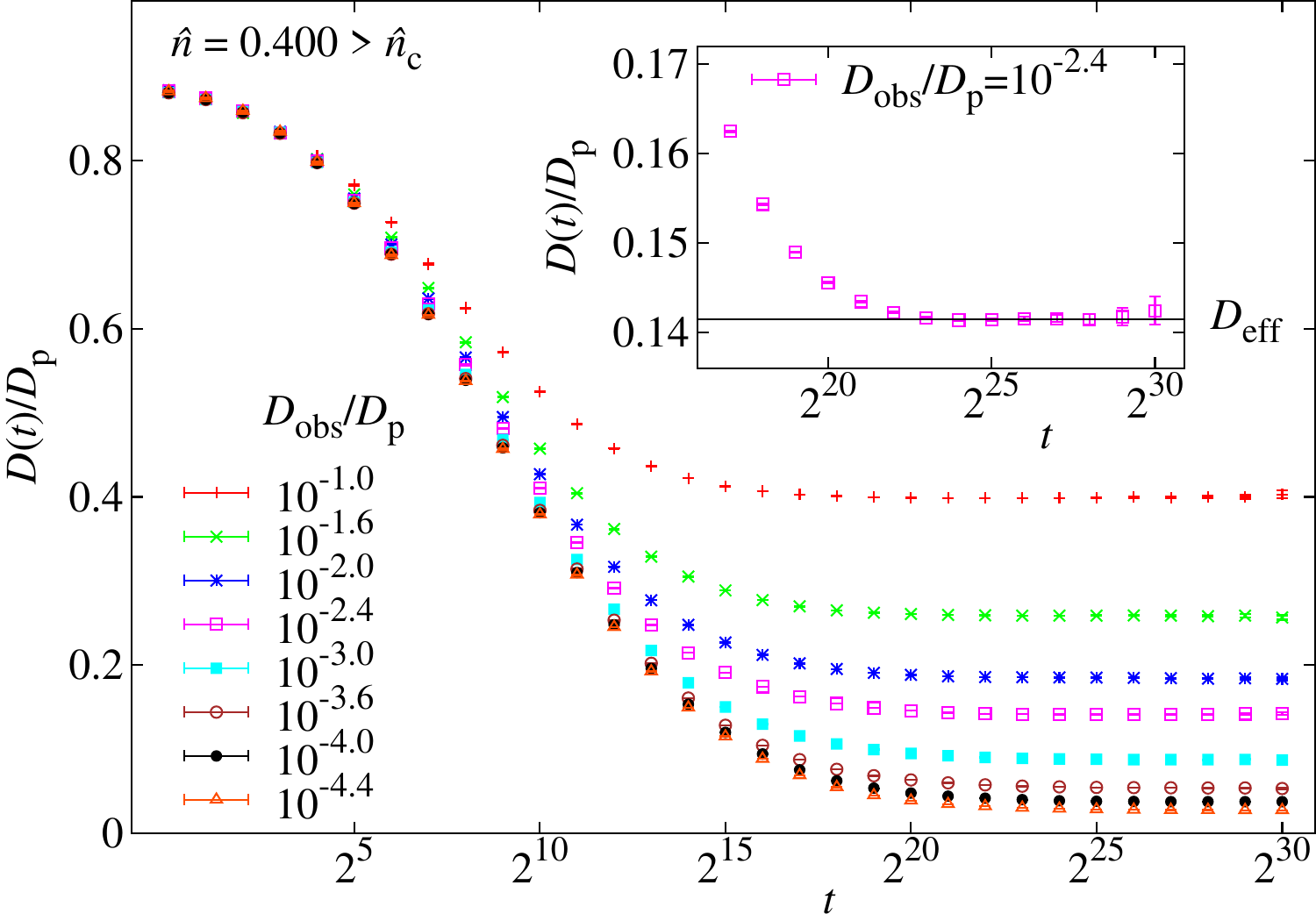}
\end{minipage}
\caption{\emph{Left:} Same as Figure~\ref{dt}, but for a subcritical
obstacle concentration ($\nhat = 0.2$). \emph{Right:}
Same as Figure~\ref{dt}, but for a supercritical
obstacle concentration ($\nhat = 0.4$). The inset shows that the 
long-time diffusive regime [constant $D(t)$] is clearly reached.
\label{fig:subsup}}
\end{figure}

Figure~\ref{dt} shows $D(t)$ for many obstacle diffusivities 
at $\nhat\approx\nhatc$. In Figure~\ref{fig:subsup}
we reproduce the same plot for a subcritical density ($\nhat = 0.2$)
and a supercritical one ($\nhat = 0.4$). The resulting plots
are qualitatively the same as Figure~\ref{dt}. Notice in particular that, 
even in the supercritical case, above the percolation point, 
the diffusive regime is eventually reached.

\section*{References}

\providecommand{\newblock}{}

\end{document}